\documentclass[a4paper]{jpconf}
\usepackage[pdftex]{graphicx}

\begin{document}
\title{Discrimination of Dark Matter Mass and Velocity Distribution by Directional Detection}

\author{Keiko I. Nagao\footnote{This talk is based on collaboration \cite{Nagao:2017yil}.}}

\address{Ridaicho 1-1, Kitaku, Okayama, 700-0005, JAPAN}

\ead{nagao@dap.ous.ac.jp}

\begin{abstract}
The velocity distribution of dark matter is supposed to be isotropic Maxwell-Boltzmann distribution in most cases, 
however, its anisotropy is 
suggested by some simulations. 
We investigate conditions to give constraints on the anisotropy with directional direct detections.
\end{abstract}

\section{Introduction}
Directional direct detection of dark matter is designed to detect both the recoil energy of dark matter-nucleus scattering and
direction of the nuclear recoil
while only the recoil energy is detected in the ordinary direct detection experiment.
Due to the directional detecting capability, it is a hopeful candidate to discriminate the velocity distribution of dark matter,
especially, its anisotropy. In this study, we investigate conditions to discriminate the Maxwell-Boltzmann velocity distribution from an anisotropic velocity distribution suggested by N-body simulation in the directional direct detection experiments.

\subsection{Anisotropy of velocity distribution}
In most cases, the velocity distribution of dark matter is assumed to be Maxwell-Boltzmann distribution. However, 
other possibilities are proposed by observations and N-body simulations. 
For example, tangential velocity distribution $f(v_\phi)$ in the Galactic rest frame consists of the isotropic component 
and anisotropic component in Eq. \cite{LNAT}):
\begin{eqnarray}
f(v_\phi) = \frac{1-r}{N(v_{0,\mathrm{iso.}})} \exp\left[-v_\phi^2/v_{0,\mathrm{iso.}}^2\right] + 
		\frac{r}{N(v_{0,\mathrm{ani.}})} \exp\left[-(v_\phi-\mu)^2/v_{0,\mathrm{ani.}}^2\right] ,
\label{eq:doublegaussian}
\end{eqnarray}
where $N(v_{0,\mathrm{iso.}})$ and $N(v_{0,\mathrm{ani.}})$ are normalization factors,  $r$ is anisotropic parameter, $v_{0,\mathrm{iso.}} = 250$ km/s, 
$v_{0,\mathrm{ani.}} = 120$ km/s and $\mu=150$ km/s. The simulation yields anisotropy $r=0.25$, while in the Maxwell-Boltzmann velocity distribution it corresponds to $r=0.00$.

\subsection{Directional direct detection}
Directional direct detection of dark matter is next-generation detection in which both the recoil energy and its direction can be detected \cite{Mayet:2016zxu}. Currently several projects are in research and development. Most of them uses gaseous detector
like CF$_4$, CS$_2$ and CHF$_3$. It has sensitivity for the spin-dependent cross section of dark matter and nucleus scattering. 
As a solid detector, we suppose NEWSdm in which nuclear emulsion is used. It has several target atoms including silver (Ag), bromine (Br) and carbon (C). Among the target atoms, we suppose F and Ag as typical light target and heavy target in the numerical simulation, respectively.

\section{Numerical analysis}
In this study, Monte-Carlo simulation of dark matter-nucleus scattering is generated as follows.
\begin{enumerate}
\item A dark matter particle is generated with velocity following the velocity distribution (\ref{eq:doublegaussian}). 
\item It scatters a target nucleus, and the recoil energy $E_R$ and its scattering angle $\theta$ in the laboratory system are obtained.
\item By repeating (i)--(ii), an event distribution in the detector is obtained.
\end{enumerate}
Analysis of the obtained distribution depends on resolution of the detector. In case that the energy resolution of the detector is not enough, only an angular distribution is obtained, while an energy-angular distribution can be analyzed in case that both energy and angular resolution are good enough. Therefore we investigate angular distribution and energy-angular distribution. 

\subsection{Anisotropy}
\begin{figure}[t!]
 \begin{minipage}{0.5\hsize}
  \begin{center}
   \includegraphics[width=50mm,  angle=270]{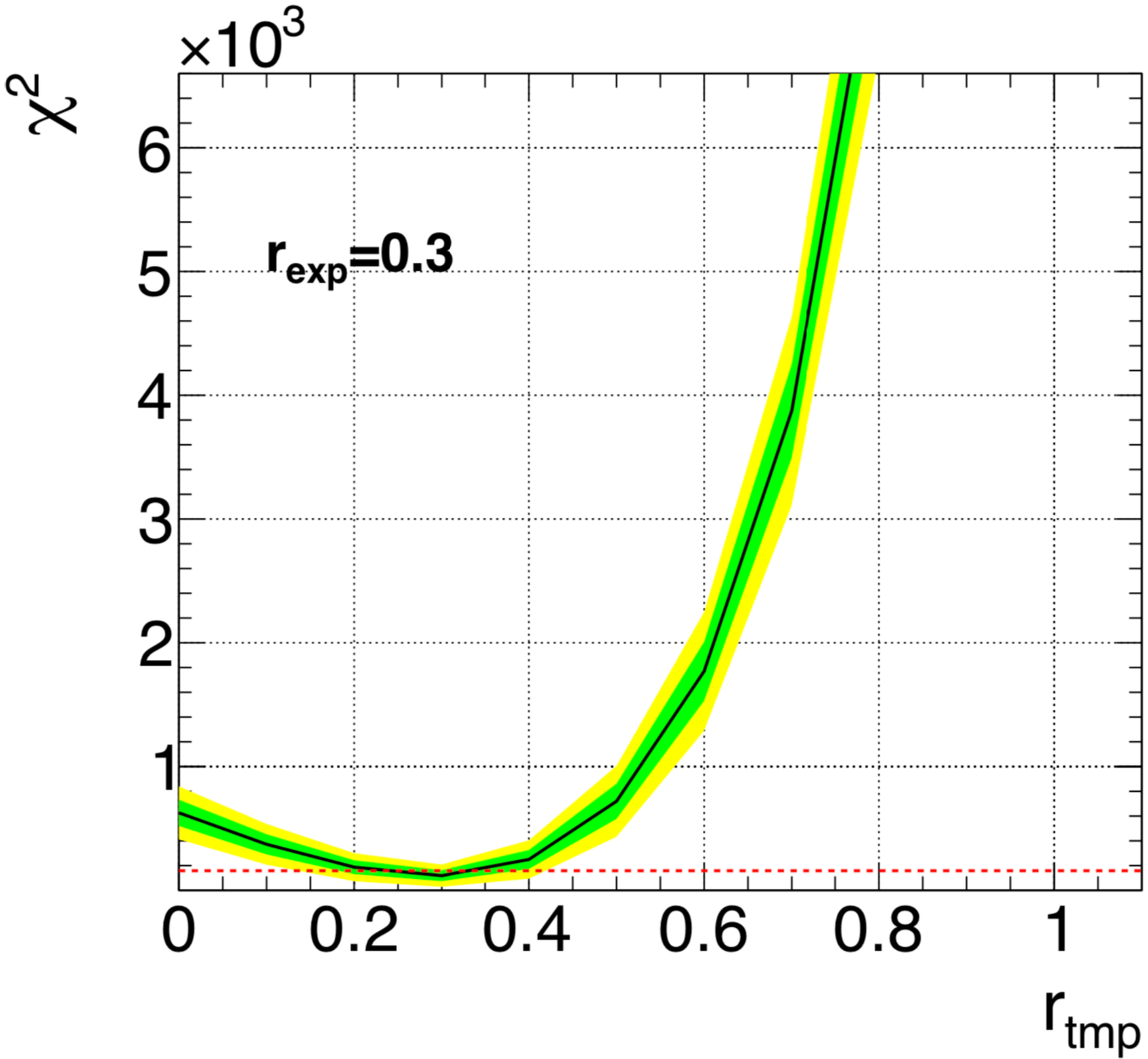}
  \end{center}
 \end{minipage}
 \begin{minipage}{0.5\hsize}
  \begin{center}
   \includegraphics[width=50mm, angle=270]{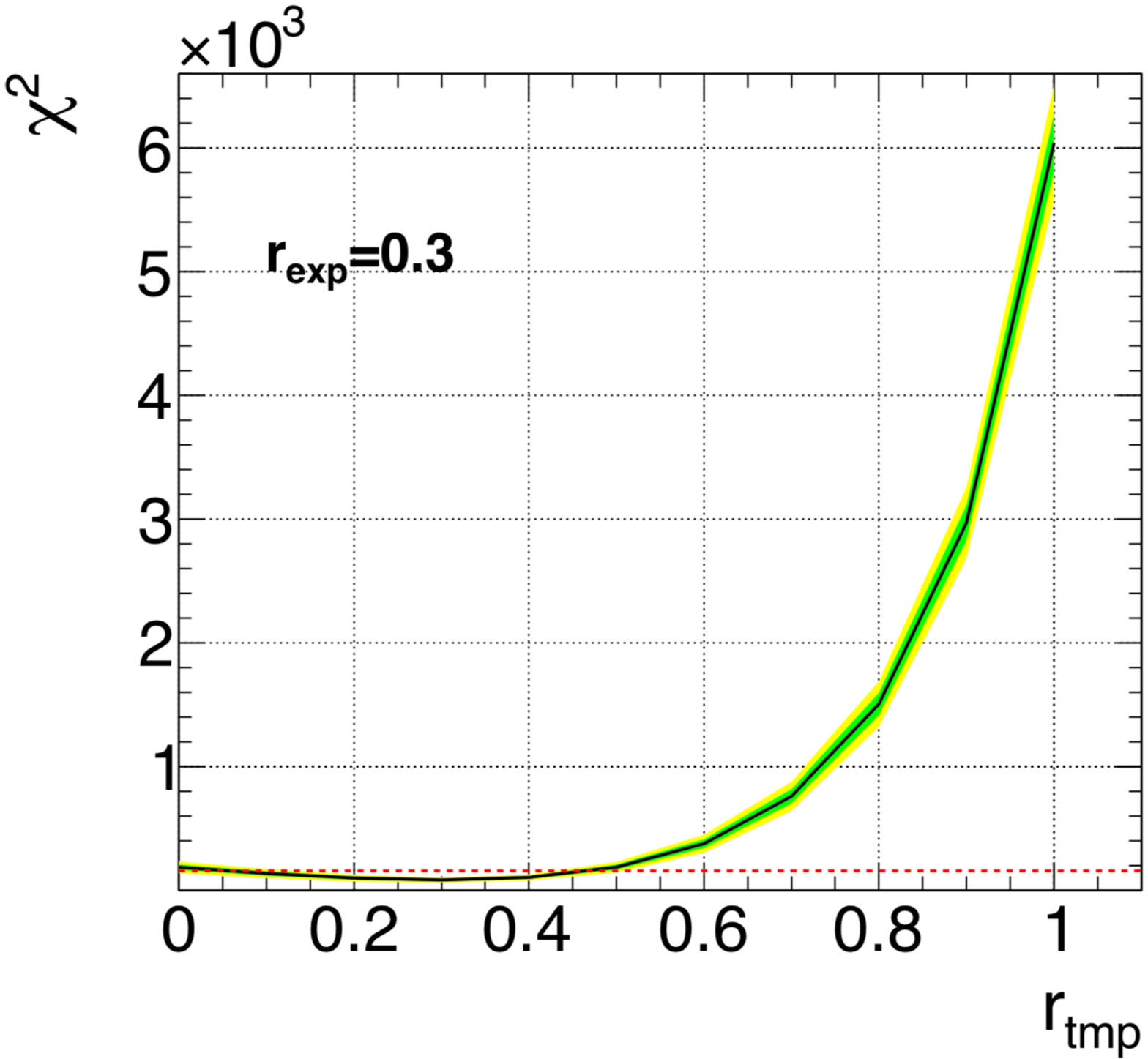}
  \end{center}
  \end{minipage}\\
 \begin{minipage}{0.5\hsize}
  \begin{center}
   \includegraphics[width=50mm, angle=270]{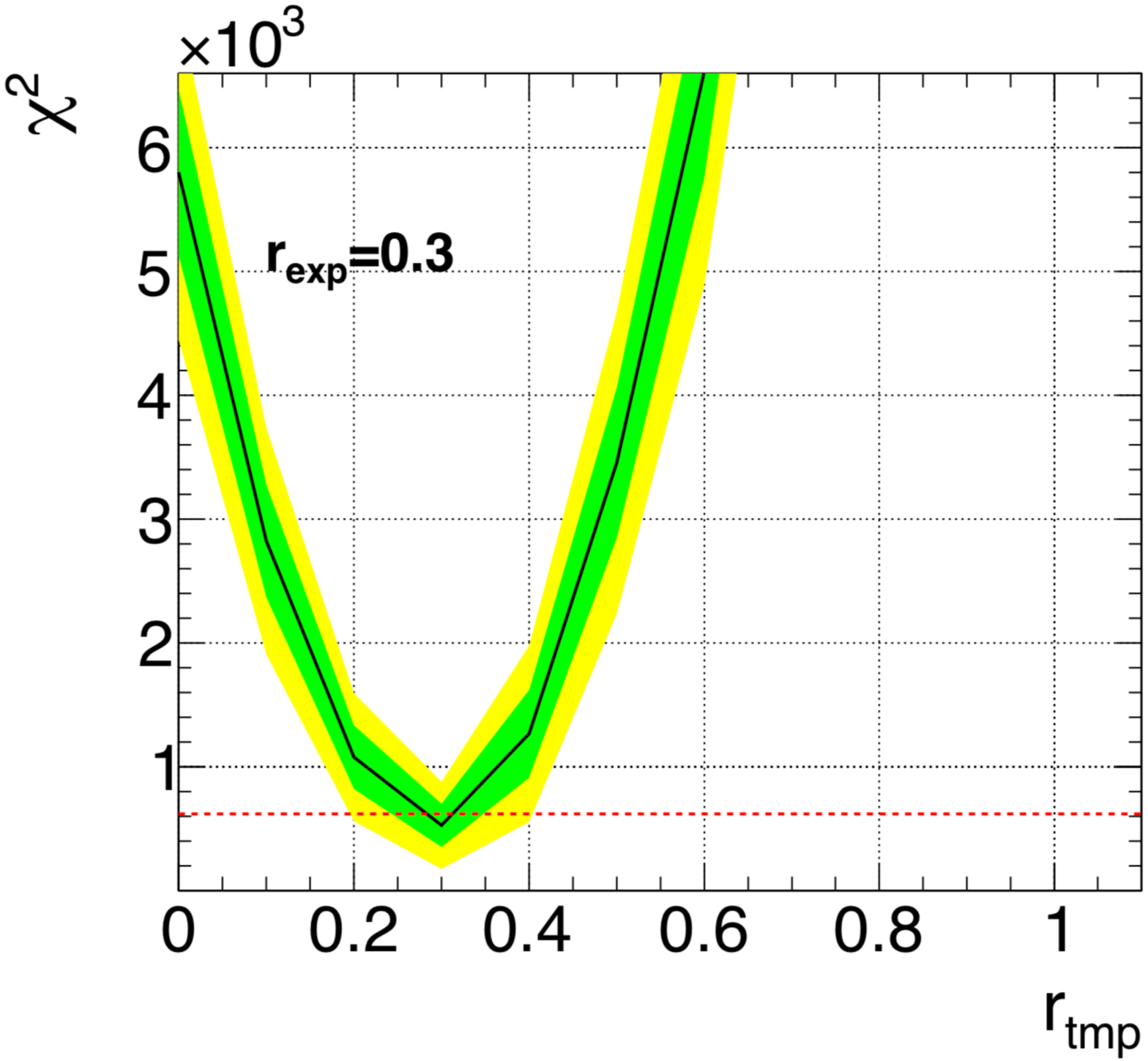}
  \end{center}
 \end{minipage}
 \begin{minipage}{0.5\hsize}
  \begin{center}
   \includegraphics[width=50mm, angle=270]{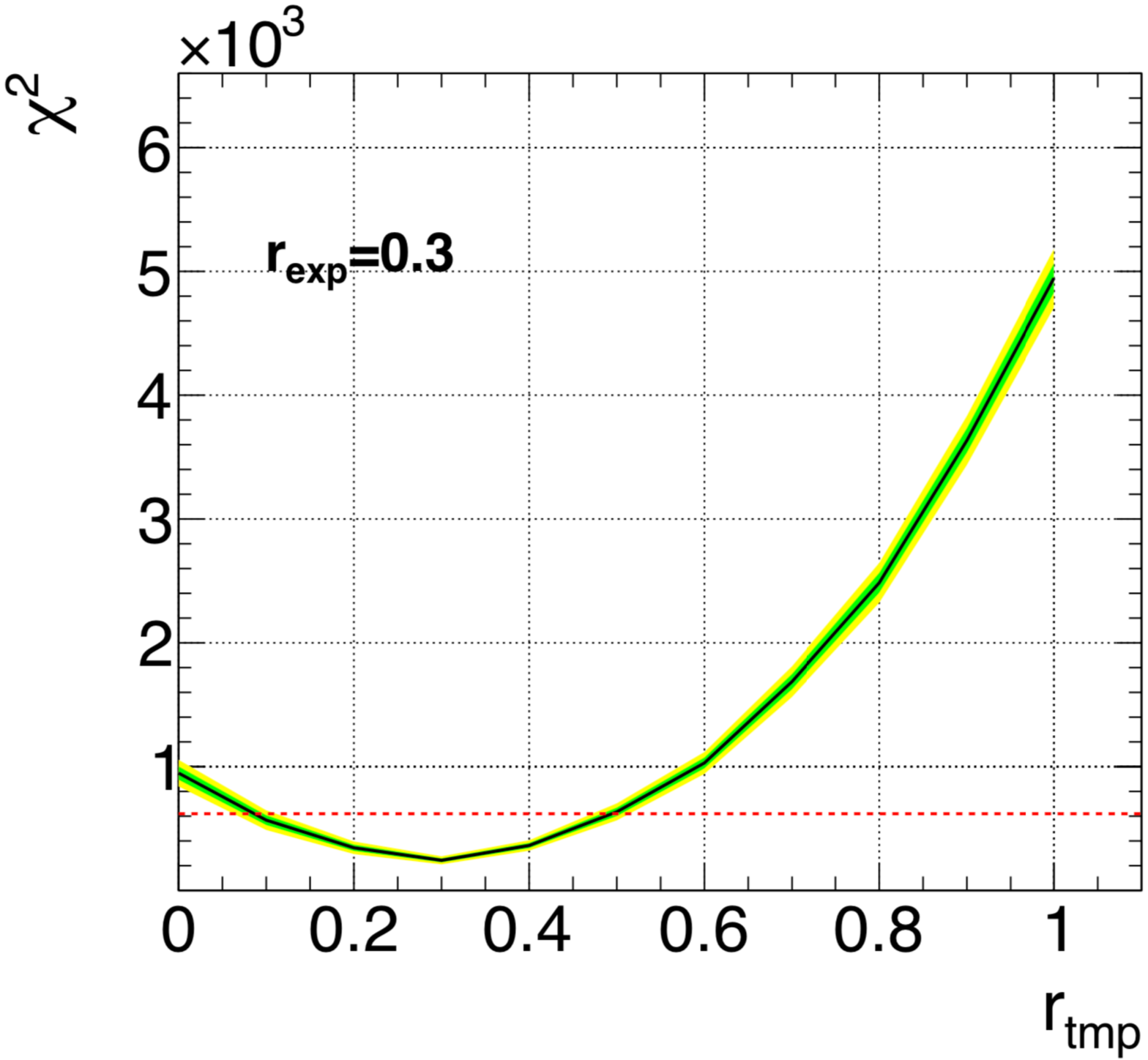}
  \end{center}
 \end{minipage}
 \caption{Chi-squared test for the angular distribution (top left) and the energy-angular distribution (top right) for target F. 
 Same results for target Ag are shown in bottom left and bottom right, respectively. Red lines represent 90\% confidence level (C.L.). Event numbers of pseudo-experiment are $5\times 10^3$(top left), $6\times 10^3$ (top right), $2\times 10^4$ (bottom left) and $6\times 10^4$ (bottom right). Dark matter mass and the energy threshold of the detector is 57 GeV and 20 keV for target F, and 324 GeV and 50 keV for target Ag, respectively.}
   \label{fig:one}
\end{figure}
If the dark matter mass is suggested by other experiments like indirect detections or collider searches,  
an energy threshold of the detector can be optimized. To investigate the possibility for the discrimination, 
we generate two kinds of data set. A data set has plenty event number such as $O(10^8)$ which corresponds to ideal distribution. 
The other data set has less event number than ideal case which corresponds to the event number in a realistic experiment. The former and the latter are hereinafter referred as template and pseudo-experimental data.
 In Figure \ref{fig:one}, results of chi-squared test between the template and pseudo-experimental data are shown. Supposing the anisotropic distribution is realized, the isotropic Maxwell-Boltzmann distribution can be rejected at 90\% C.L. with $\sim O(10^{3-4})$ for target F and $\sim O(10^{4-5})$ for target Ag. 

\subsection{Dark matter mass and anisotropy}
\begin{figure}[htbp]
 \begin{minipage}{0.33\hsize}
  \begin{center}
   \includegraphics[width=40mm, angle=270]{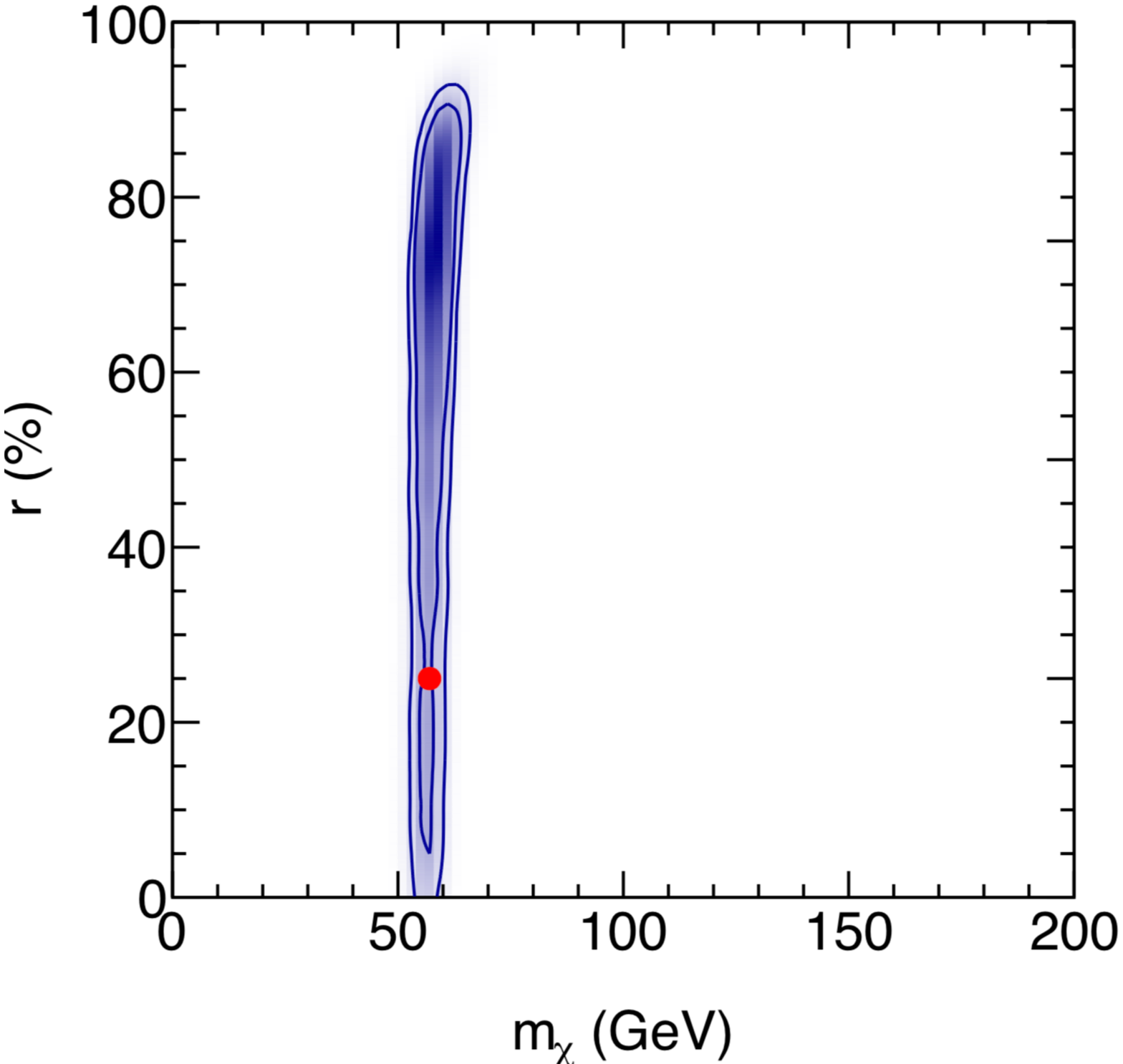}
  \end{center}
 \end{minipage}
 \begin{minipage}{0.33\hsize}
 \begin{center}
  \includegraphics[width=40mm, angle=270]{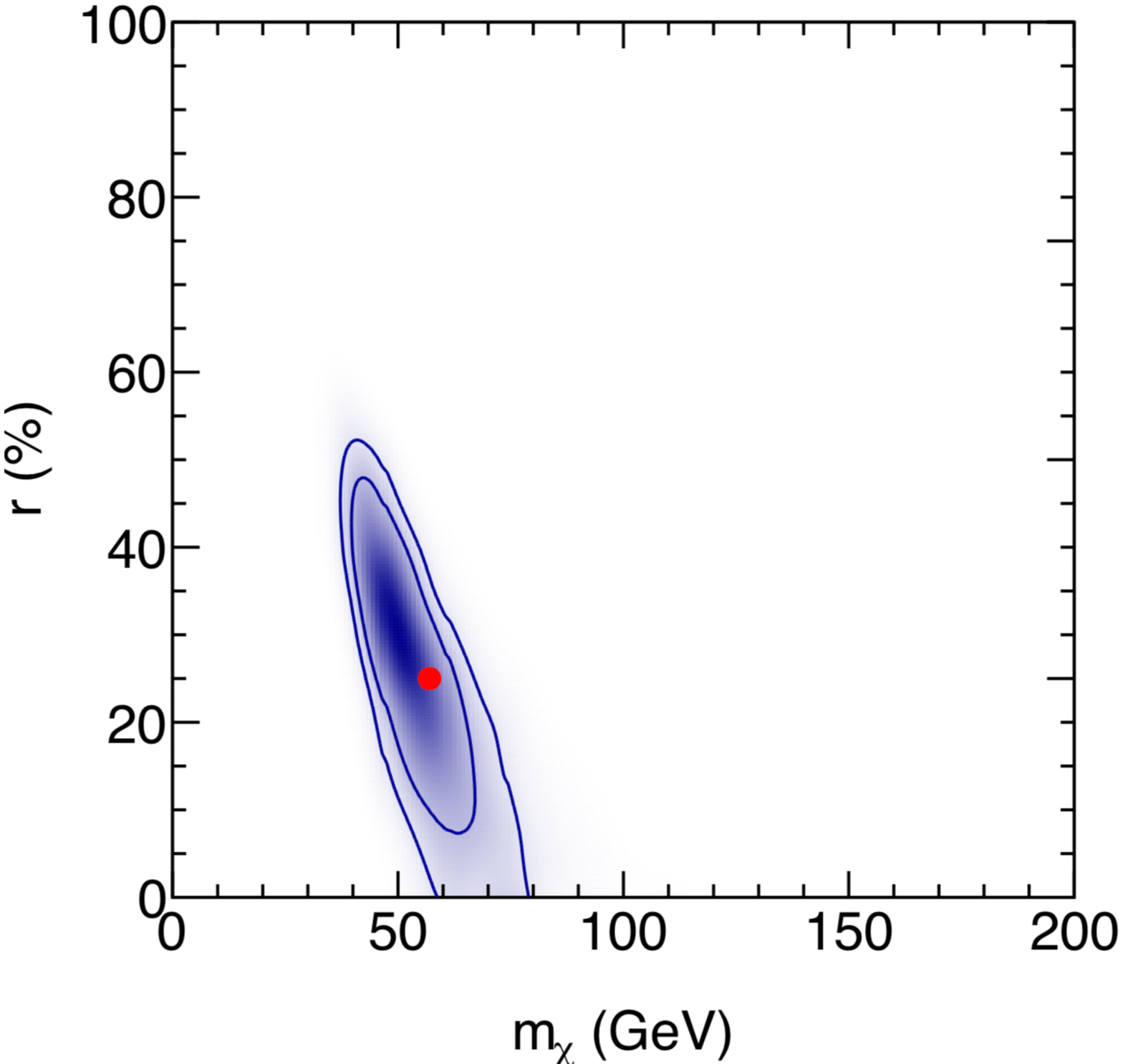}
 \end{center}
 \end{minipage}
 \begin{minipage}{0.33\hsize}
 \begin{center}
  \includegraphics[width=40mm, angle=270]{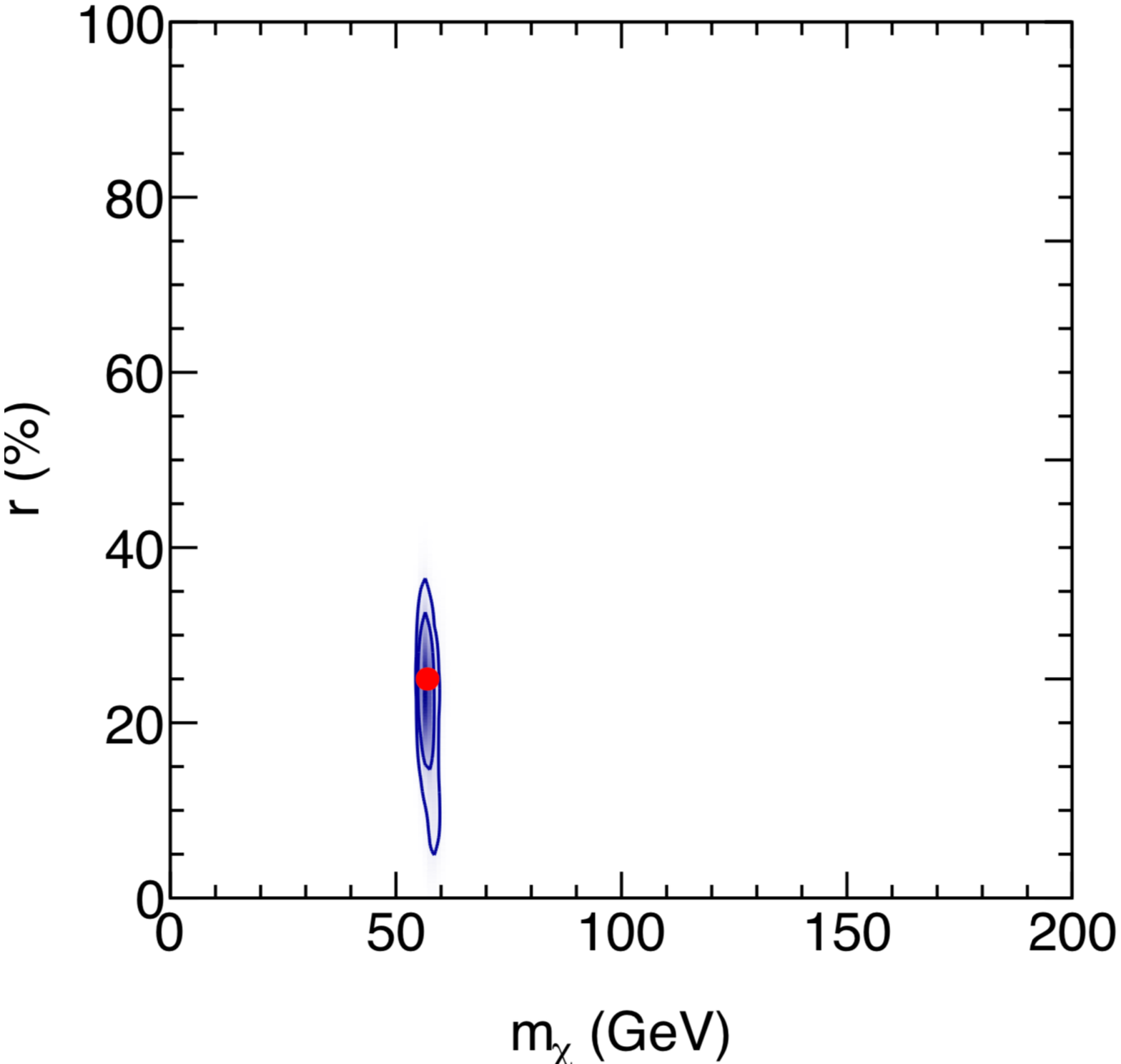}
 \end{center}
 \end{minipage}
 \vspace{-0.5cm}\\
 \begin{minipage}{0.33\hsize}
  \begin{center}
   \includegraphics[width=40mm, angle=270]{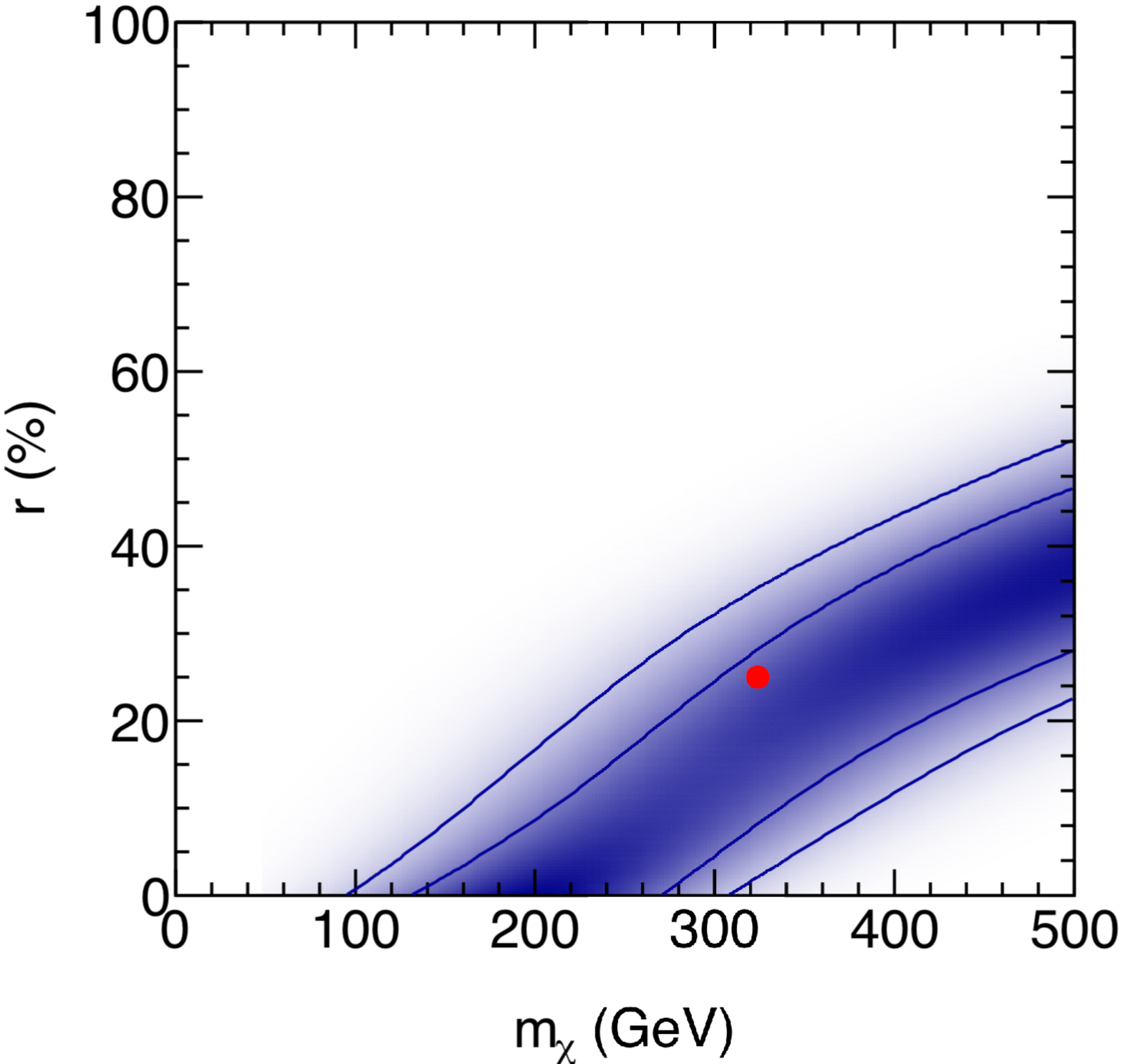}
  \end{center}
 \end{minipage}
 \begin{minipage}{0.33\hsize}
 \begin{center}
  \includegraphics[width=40mm, angle=270]{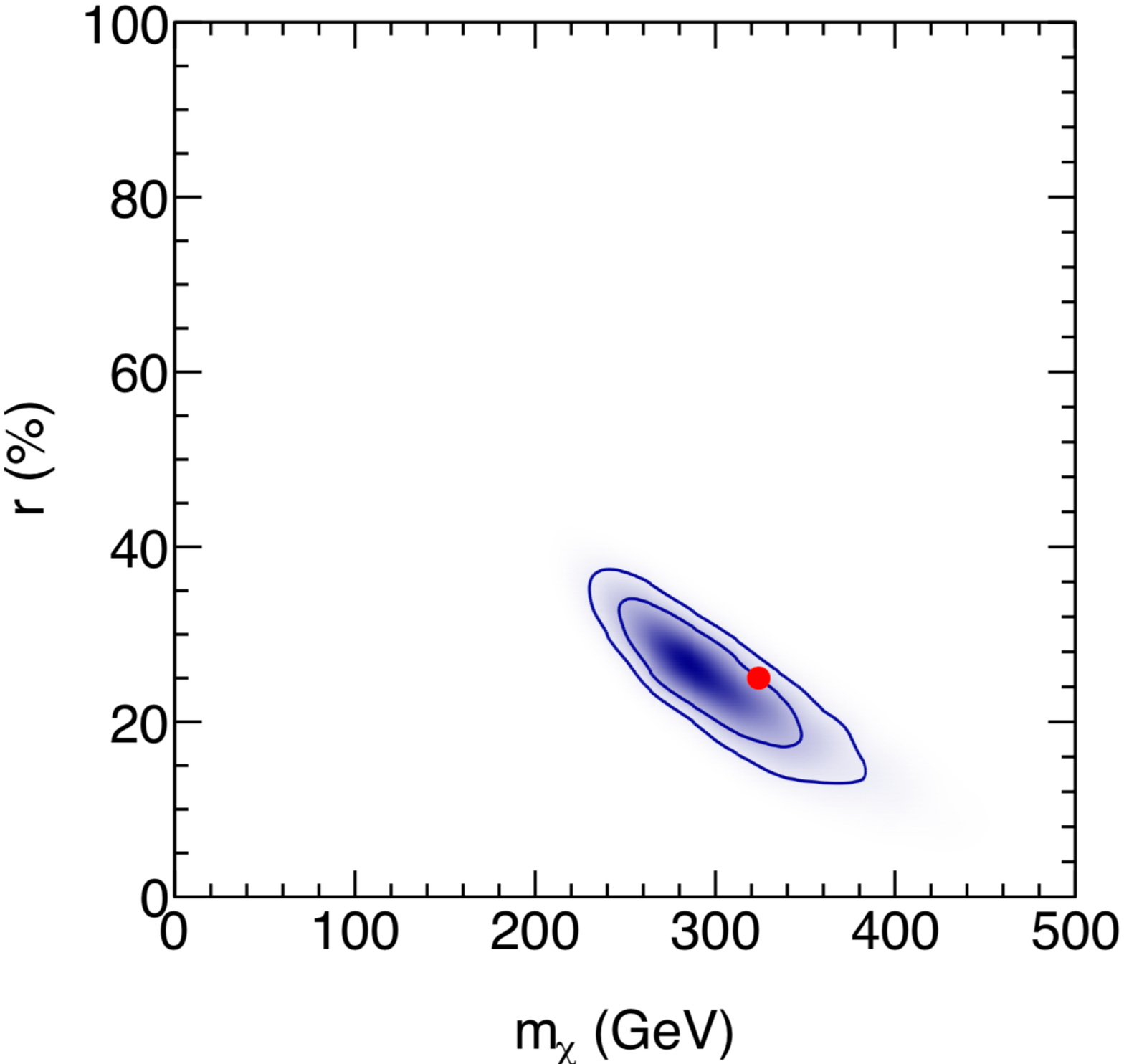}
 \end{center}
 \end{minipage}
 \begin{minipage}{0.33\hsize}
 \begin{center}
  \includegraphics[width=40mm, angle=270]{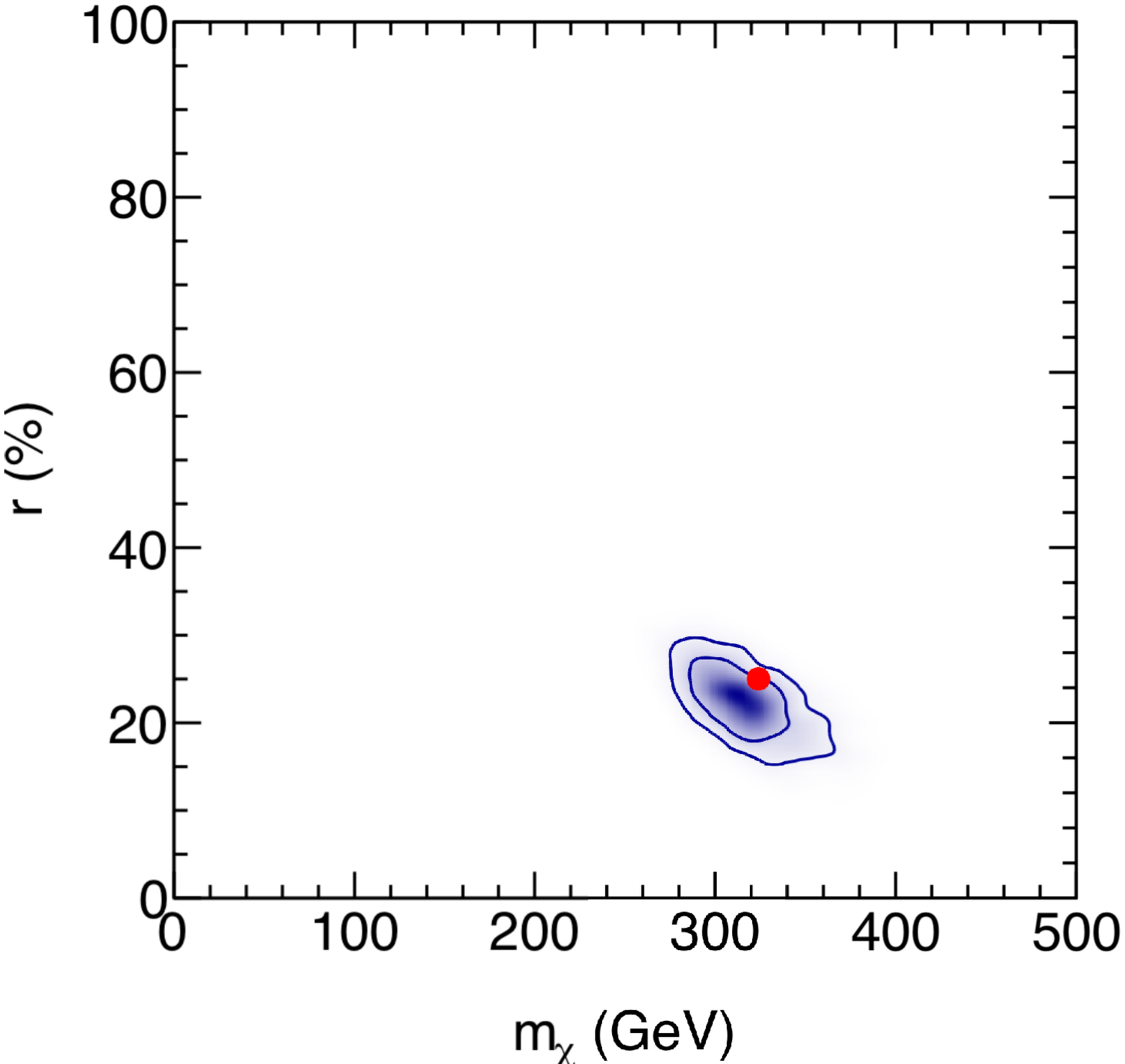}
 \end{center}
 \end{minipage}
   \label{fig:two}
   \caption{Posterior probability obtained by likelihood analysis for dark matter mass $m_\chi$ and anisotropy $r$. 
   Top and bottom rows correspond to 
   the case of target F and Ag, respectively. Left, middle and right columns correspond to analysis of the energy distribution, angular distribution, and energy-angular distribution, respectively. Inner and outer contours show 68\% and 90\% C.L, respectively. Red dots show sample points assumed in the simulation. }
\end{figure}

Even if the dark matter mass is unknown, constraints for mass and anisotropy can be obtained. In Figure \ref{fig:two}, 
likelihood analysis for the mass and the anisotropy is shown. With both energy and directional information, 
the most faithful constraint can be obtained as seen from the figures.

\section{Conclusion}
Anisotropy of the dark matter velocity distribution can be constrained in the directional direct detection. 
If dark matter mass is known,
$\sim O(10^{3-4})$ for light target atom and $\sim O(10^{4-5})$ for heavy target atom are required to give the constraint. 
They correspond to dark matter-nucleus cross section $\sigma_n \simeq 10^{-30}$ cm$^2$ with exposure kg$^{-1}$day$^{-1}$ for F and 
$\sigma_n \simeq 10^{-28}$ cm$^2$ for Ag.


\section*{References}
\begin{thebibliography}{9}
\bibitem{Nagao:2017yil} 
  K.~I.~Nagao, R.~Yakabe, T.~Naka and K.~Miuchi,
  arXiv:1707.05523 [hep-ph].
  \bibitem{LNAT}
 F.~S.~Ling, E.~Nezri, E.~Athanassoula and R.~Teyssier,
  JCAP {\bf 1002}, 012 (2010).

\bibitem{Mayet:2016zxu} 
  F.~Mayet {\it et al.},
  Phys.\ Rept.\  {\bf 627}, 1 (2016)
\end{thebibliography}
\end{document}